\documentclass[twocolumn,prl,aps,superbib,tightenlines,floatfix]{revtex4}
\usepackage{amsfonts,amsmath,amssymb,amsthm,bm}
\usepackage{graphicx}
\usepackage[dvipsnames,usenames]{color}
\begin{document}
\bibliographystyle{apsrev}
\title{Thermoelectricity of Molecular tunnel Junctions}
\author{Yu-Shen Liu }
\author{Yu-Chang Chen}
\email{yuchangchen@mail.nctu.edu.tw} \affiliation{Department of
Electrophysics, National Chiao Tung University, 1001 Ta Hsueh Road,
Hsinchu 30010, Taiwan }
\begin{abstract}

A first-principles approach is presented for the thermoelectricity
in molecular junctions formed by a single molecule contact. The
study investigates the Seebeck coefficient considering the
source-drain electrodes with distinct temperatures and chemical
potentials in a three-terminal geometry junction. We compare the
Seebeck coefficient in the amino-substituted and unsubstituted
butanethiol junction and observe interesting thermoelectric
properties in the amino-substituted junction. Due to the novel
states around the Fermi levels introduced by the amino-substitution,
the Seebeck coefficient could be easily modulated by using gate
voltages and biases. When the temperature in one of the electrodes
is fixed, the Seebeck coefficient varies significantly with the
temperature in the other electrode, and such dependence could be
modulated by varying the gate voltages. As the biases increase, richer
features in the Seebeck coefficient are observed, which are closely
related to the transmission functions in the vicinity of the left and
right Fermi levels.

\end{abstract}
 \maketitle

Building electronic circuits from molecules is an inspiring idea
\cite{Aviram,Ahn,Lindsay,Tao1}. Much attention has been devoted to
investigating the various transport properties that might be
applicable in developing new forms of electronic and
energy-conversion devices, such as electron transfer
\cite{Huynh,Nitzan1}, shot noise \cite{Shi}, heat transport
\cite{Nitzan2,Galperin1}, negative differential resistance
\cite{ChenJ}, and gate-controlled effects \cite{Di Ventra1}.
Recently, topics on thermo-related transport, such as local
heating \cite{Galperin1,Huang1,Huang2} and thermal transport
\cite{Prasher}, have emerged as new subfields in molecular
electronics. Another important thermo-related property in the
molecular tunnel junction (m-M-m) is thermoelectricity \cite{
Dubi,Paulsson,Pauly,Segal,Zheng,Ludoph,Reddy,Galperin2,Baheti}.
The Seebeck coefficient, which is related not only to the
magnitude but also to the slope of the transmission function
in the vicinity of Fermi levels, can provide more information
than current-voltage characteristics. The study of thermoelectricity
is of key importance in the design of novel thermo-related electronic
and nanoscale energy conversion devices.

Recent experimental measurements of the Seebeck coefficient for
molecular tunnel junctions were conducted at zero bias
\cite{Reddy,Baheti}. In such cases, the system can be described by
only a single Fermi level. Similarly, the Seebeck coefficient for
bulk material is also described by a single Fermi level.
Nevertheless, molecular tunnel junctions consist of two electrodes
as independent electron and heat reservoirs. Thus, it is worthwhile
extending the investigation of the Seebeck coefficients to a system
with distinct temperatures ($T_{L(R)}$) and chemical potentials
($\mu_{L(R)}$) in the left (right) electrode. In this
letter, we present a theory for two distinct Fermi levels in molecular
tunnel junction combining a first-principles approach for the Seebeck
coefficient in the two- and three-terminal junctions in nonlinear
regime. As an example, we systematically
investigate the dependence of the Seebeck coefficient on the
source-drain biases, gate voltages, and temperatures in the metal
electrodes before and after amino-substitution in the butanethiol
molecular junction. The Seebeck coefficient may provide further
insights into the physical properties of molecular tunnel
junctions. For example, whether the Fermi energy is closer to the
lowest unoccupied molecular orbital (LUMO) or the highest occupied
molecular orbital (HOMO) may be locally probed via thermoelectricity
measurements \cite{Paulsson}. Interesting features in the Seebeck
coefficient are observed in the amino-substituted butanethiol
junction because of the dramatic change in the transmission functions
by amino-substitution in the vicinity of the left and right Fermi
levels.

Alkanethiol [CH$_{3}$(CH$_{2})_{n-1}$SH, denoted as C$_{n}$]-related
molecules are a good representation of reproducible junctions that
can be fabricated \cite{Ahn,Lindsay,Tao2}. It has been established
that non-resonant tunneling is the main conduction mechanism inasmuch as the
Fermi levels of the two electrodes lie within the large HOMO-LUMO
gap. However, functional group substitution may have significant
effects on the electronic structures of alkanethiols. New states around
the Fermi levels are produced when -NH$_{2}$ is substituted
for -H in bridging butanethiol (C$_{4})$. The response of these states to
external biases depends on the polarity and it leads to asymmetric
current-voltage characteristics \cite{Ma}.
Because of the dramatic change in the transmission function in the
vicinity of the Fermi levels, the novel characteristics
of the Seebeck coefficient are observed in the amino-substituted
junction. Consequently, the amino-substitution
significantly affects the Seebeck coefficient. For example, the
Seebeck coefficient can change signs by applying gate voltages and
biases in the amino-substituted junction but not in the unsubstituted
system. The influence of different temperatures between the two
electrodes on the Seebeck coefficient, controllable by the gate
voltages, is significant in the amino-substituted junction. The
results suggest that the thermoelectric molecule devices, such as
a molecular thermometer, are possible in the future.

\begin{figure}
\includegraphics[width=8cm]{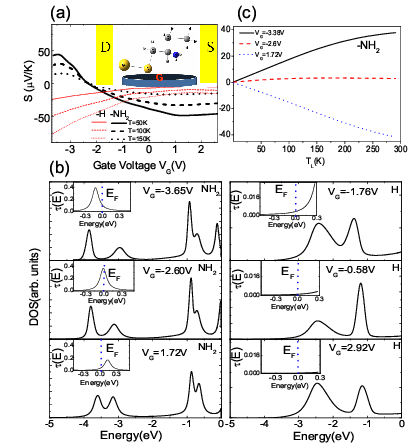}
\caption{(color online) The Seebeck coefficient S in a
three-terminal geometry with $V_{SD}=0.01V$: \textbf{(a)} The S
versus $V_G$ where $T_{L}=T_{R}=T$ for the amino-substituted
[black (thick) lines] and unsubstituted [red (thin) lines] butanethiol for
$T=50$~K (solid line), $T=100$~K (dashed line), and $T=150$~K (dotted
line). The inset shows the schematic of the three-terminal junction.
The gate field is applied in a direction perpendicular to direction
of charge transport. \textbf{(b)} The density of states and the
transmission function (inset): the left pannels for the
amino-substituted butanethiol junction at $V_G=-3.65,-2.60,$ and
$1.72$~V; the right pannels for the unsubstituted butanethiol
junction at $V_G =-1.76,-0.58,$ and $2.92$~V. \textbf{(c)} The S
versus $T_L$ for an amino-substituted butanethiol junction for
$V_G=-3.38,~-2.60,$ and $1.72$~V, where $T_R=0$~K.} \label{Fig1}
\end{figure}

Let us start by considering a single molecule sandwiched between two
bulk electrodes applied with a certain source-drain bias. The Fermi
level in the left/right electrodes
is determined by filling the conduction band with
the valence electrons in the bulk Au electrode described by the
jellium model ($r_s\approx 3)$. The gate voltage is introduced as a
capacitor composed of two parallel circular charged disks separated
by a certain distance from each other \cite{Di Ventra1,Di Ventra2}.
The axis of the capacitor is perpendicular to the transport
direction. One plate is placed close to the molecule, while the
other plate, placed far away from the molecule, is set to be the
zero reference energy [Inset in Figure~\ref{Fig1}(a)]. Using the
second-quantization field-operator technique with the effective
single-particle wave functions calculated self-consistently
in density functional theory, the current is given by \cite{Chen}:
\begin{eqnarray}
I=\frac{1}{\pi }\int {dE\left[ {f_E^R (\mu _R ,T_R
)\tau^R(E)-f_E^L (\mu _L ,T_L )\tau^L(E)}\right]}, \label{current}%
\end{eqnarray}
where the transmission function of electron with energy E
incident from the left (right) electrode is:
\begin{eqnarray}
\tau ^{L(R)}(E)=\pm i\pi \int {d{\rm {\bf R}}\int {d{\rm {\bf
K}}_{\vert \vert } } } I_{EE}^{LL(RR)} ({\rm {\bf r}},{\rm {\bf
K}}_{\vert \vert }), \label{trans}%
\end{eqnarray}
where $I_{E{E}'}^{ij}=[\Psi _E^i ]^\ast \nabla \Psi _{{E}'}^j
-\nabla [\Psi _E^i ]^\ast \Psi _{{E}'}^j$, and $i,~j=L,~R$. $\Psi
_E^{L(R)}({\rm{\bf r},\rm {\bf K}}_{\vert \vert })$ is the
single-particle wave function (detailed theory can be found in
Ref.~\cite{Chen,Lang}) incident from the left (right)
electrode with energy $E$ and component of the momentum ${\rm {\bf
K}}_{\vert \vert }$ parallel to the electrode surface, and $d{\rm
{\bf R}}$ represents an element of the electrode surface. The
stationary wave function  $\Psi _E^{L(R)}({\rm{\bf r},\rm {\bf
K}}_{\vert \vert })$ can be calculated by solving the
Lippmann-Schwinger equation iteratively to self-consistency
\cite{Lang,Di Ventra3,Yang}. The exchange-correlation potential is
included in density-functional formalism by using the local-density
approximation \cite{Hoh}. Once the single-particle wave functions
are calculated self-consistently, the transmission function of
electron with energy E can be calculated using Eq.~(\ref{trans}).
We assume that the left (right) electrode serves as the electron
and thermal reservoir with the electron population described by the
Fermi-Dirac distribution function, $f_{E}^{L(R)}=1/\left( \exp
\left( \left( E-\mu _{L(R)}\right) /k_{B}T_{L(R)}\right) +1\right)$,
where $\mu_{L(R)} $ and $T_{L(R)} $ are the chemical potential and the
temperature in the left (right) electrode, respectively, and $k_B$
is the Boltzmann constant.

This research considers the extra current induced by an additional
infinitesimal temperature ($\Delta T$) and voltage ($\Delta V$)
distributed symmetrically across the molecular junction:
\begin{widetext}
\begin{eqnarray}
\Delta I &=&I(\mu _{L},T_{L}+\frac{\Delta T}{2};\mu _{R},T_{R}-\frac{\Delta T%
}{2})+I(\mu _{L}+\frac{e\Delta V}{2},T_{L};\mu _{R}-\frac{e\Delta V}{2}%
,T_{R})-2I(\mu _{L},T_{L};\mu _{R},T_{R}), \label{J1}
\end{eqnarray}
\end{widetext}

The Seebeck coefficient (defined as $ S=\frac{\Delta V}{\Delta T}$)
is obtained by letting $\Delta I=0$. We expand the Fermi-Dirac
distribution function to the first order in $\Delta T$ and $\Delta
V$ and obtain

\begin{eqnarray}
S=-\frac{1}{e}\frac{\frac{K_{1}^{L}}{T_{L}}+\frac{K_{1}^{R}}{T_{R}}}{%
K_{0}^{L}+K_{0}^{R}} ,\label{S1}
\end{eqnarray}
where
\begin{eqnarray}
K_{n}^{L(R)}=-\int dE\left( E-\mu _{L(R)}\right)
^{n}\frac{\partial f_{E}^{L(R)}}{\partial E}\tau (E) ,\label{S1P}
\end{eqnarray}
and $\tau (E)=\tau ^R(E)=\tau ^L(E)$, a direct consequence of the
time-reversal symmetry.

The research explores the dependence of the Seebeck
coefficient on the gate voltages, temperatures in the electrodes,
and the source-drain biases in both the linear and nonlinear
response regimes by applying Eq.~(\ref{S1}). In the low-temperature
regime where the higher order terms in the
temperature are disregarded, Eq.~(\ref{S1}) can be simplified
using the Sommerfeld expansion \cite{Zheng,Ludoph,Wang}:
\begin{eqnarray}
S=-\frac{\pi ^2k_B^2 }{3e}\frac{T_L \textstyle{{\partial \tau (E)}
\over {\partial E}}\vert _{E=\mu _L } +T_R \textstyle{{\partial \tau
(E)} \over {\partial E}}\vert _{E=\mu _R } }{\tau (\mu _L )+\tau
(\mu _R )}\quad , \label{S2}
\end{eqnarray}
where the Seebeck coefficient is closely related to the transmission
function in the vicinity of the left and right Fermi levels.

As the first step in our analysis, we study the
Seebeck coefficient in a three-terminal geometry in the linear
response regime ($V_{SD}=0.01$~V and $\mu _L \approx \mu _R \approx
E_F )$, where both electrodes have the same temperatures
($T_L=T_R=T$). In this case,  the Seebeck coefficient can be
simplified as $
 S=-\frac{1}{eT}\frac{\int {(E-E_F
)\textstyle{{\partial f_E } \over {\partial E}}\tau (E)dE} }{\int
{\textstyle{{\partial f_E } \over {\partial E}}\tau (E)dE} }$. In
the low temperature regime, the Seebeck coefficient can be further
simplified using the Sommerfeld expansion as $S=-\frac{\pi ^2k_B^2
T}{3e}\frac{\partial ln\tau (E)}{\partial E}\vert _{E=E_F } $
\cite{Galperin1,Paulsson,Galperin2}. This equation has been applied
to the study of several atomic and molecular systems \cite{Zheng,Reddy}.
The Seebeck coefficient as a function of the gate voltage for
various temperatures in the amino-substituted and unsubstituted
butanethiol junction is presented in Fig.~\ref{Fig1}(a). The results
show that the characteristics of the Seebeck coefficient are
sensitive to the gate voltages in the amino-substituted butanethiol
junction. The most striking feature is that the molecular transistor
can be converted from n-type to p-type by applying the gate
voltages. The Seebeck coefficient is close to zero at $V_G \approx
-2.6$~V. As the gate voltage further decreases, the sign of the
Seebeck coefficient becomes positive (p-type). For the butanethiol
molecular junction, the characteristic of the carrier remains n-type
all the time because the sign of the Seebeck coefficient is
negative.

To arrive at the physical reason why the gate voltage can efficiently
modulate the Seebeck coefficient, the DOSs (transmission
functions) are plotted as a function of energy for the various
gate voltages in Fig.~\ref{Fig1}(b) (Inset of Fig.~\ref{Fig1}(b)).
We observe that the positive (negative) gate voltage shifts the LUMO
peak towards higher (lower) energies. At $V_G=-2.6$~V, the peak
position of the LUMO and transmission function align with the
Fermi levels, implying that ${\partial \ln \tau (E)} \mathord{\left/
{\vphantom {{\partial \ln \tau (E)} {\partial E}}} \right.
\kern-\nulldelimiterspace} {\partial E}\vert _{E=E_F } \approx 0$.
Hence, the Seebeck coefficient is close to zero at the gate voltage
around $V_G =-2.6$~V. When the gate voltage is tuned at around $V_G
=1.72$~V, the Seebeck coefficient is negative because ${\partial \ln
\tau (E)} \mathord{\left/ {\vphantom {{\partial \ln \tau (E)}
{\partial E}}} \right. \kern-\nulldelimiterspace} {\partial E}\vert
_{E=E_F } >0$. Conversely, the Seebeck coefficient is positive because
${\partial \ln \tau (E)} \mathord{\left/ {\vphantom {{\partial
\ln \tau (E)} {\partial E}}} \right. \kern-\nulldelimiterspace}
{\partial E}\vert _{E=E_F } <0$ at $V_G =-3.65$~V. Thus, the
characteristic of the carrier type of a certain molecular junction
can be converted from n-type (closer to LUMO) to p-type (closer to
HOMO) by tuning the gate voltage. In the unsubstituted butanethiol
junction as shown in the right plane of Fig.~\ref{Fig1}(b), it is
observed that the electron transmission function is always small
because the location of the Fermi levels lies within the large
HOMO-LUMO gap. We also note that the Seebeck coefficient has a
negative value due to
 ${\partial \ln \tau (E)} \mathord{\left/
{\vphantom {{\partial \ln \tau (E)} {\partial E}}} \right.
\kern-\nulldelimiterspace} {\partial E}\vert _{E=E_F }>0$. When the
gate voltages are further decreased, the absolute value of the
Seebeck coefficient becomes bigger even in the unsubstituted
butanethiol junction because the negative gate voltage shifts the
LUMO peak towards the lower energy region.

\begin{figure}
\includegraphics[width=8cm]{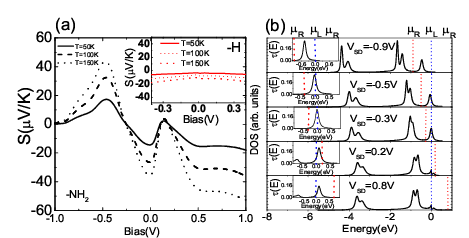}
\caption{(color online) (a)The Seebeck coefficient S as a function
of source-drain biases in a two-terminal geometry for
amino-substituted (the main graph) and unsubstituted (the inset in
the upper right corner) butanethiol for $T=50$~K (solid lines), $T=
100$~K (dashed lines), and $T=150$~K (dotted lines). (b) The density
of states and the transmission function (inset) for various
source-drain biases ($V_{SD}=-0.9$, $-0.5$, $-0.3$, $0.2$,
and $0.8$~V) in the amino-substituted butanethiol junction.}
\label{Fig2}
\end{figure}

The Seebeck coefficient is relevant to the temperatures of the
electrodes. This property may be applied to the design of a
molecular thermometer. To show this, we investigate
the Seebeck coefficient of the amino-substituted butanethiol
junction at $V_{SD}=0.01$~V as a function of temperature of the left
electrode ($T_L)$ while keeping $T_R =0$~K as shown in
Fig.~\ref{Fig1}(c). The results show that the dependence of the
Seebeck coefficient on $T_L$ is linear at low temperatures. In this
regime, the Seebeck coefficient can be well described by
Eq.~(\ref{S2}), where $T_R =0$~K. As the temperature $T_{L}$ becomes
large, the approximation of Eq.~(\ref{S2}) turns out to be
inappropriate, and the Seebeck coefficient shows nonlinear behavior.
We further observe that the sensitivity of the Seebeck coefficient
versus $T_L$ can be amplified by applying gate voltages. At $V_G
=-2.6$~V, the Seebeck coefficient has a very small value insensitive
to $T_L$ because the peak of the transmission function lies between two
Fermi levels. When the gate voltage is tuned to $V_G=-3.38$~V, the Seebeck
coefficient can be enhanced to around $38~\mu \mbox{V/K}$ at
$T_L=300$~K. The other interesting phenomenon observed is the
possibility to change the sign of the Seebeck coefficient by
applying the gate voltage. When the gate voltage is tuned to $V_G
=1.72$~V, the Seebeck coefficient becomes around $-42~\mu
\mbox{V/K}$ at $T_L =300$~K. The results show that the
amino-substituted butanethiol may be an effective thermoelectric
material applicable to the design of molecular thermoelectric
devices such as a thermometer.

The project further investigates the Seebeck coefficient in the
nonlinear regime. The Seebeck coefficient of the amino-substituted
(unsubstituted) butanethiol molecular junctions as a function of
$V_{SD}$ is plotted in Fig.~\ref{Fig2}(b) (Inset of
Fig.~\ref{Fig2}(b)). At large $V_{SD}$, the difference between the
left and right chemical potentials  becomes  significant. Thus, the
transmission functions in the vicinity of both the left
and right Fermi levels have important contribution to the Seebeck
coefficient. In Fig.~\ref{Fig2}(b) we plot the DOSs and transmission
functions as functions of the energy for various biases.
For $V_{SD}>0$, the states between the left and right
Fermi levels are developed into a resonant peak similar to what is
found in the elongated silicon point contact \cite{Di Ventra4}.
At large $V_{SD}$, the transmission functions around the left and
right Fermi levels are equally important to the Seebeck coefficient
(see Eq. (\ref{S2})). When $T_L =T_R =T$, we explain the Seebeck
coefficient in Fig.~\ref{Fig2}(a) by considering the source-drain
bias $V_{SD}$ at $0.8,~-0.3$, and $-0.5$~V, respectively. For these
biases, the contribution to the Seebeck coefficient is dominated by
the transmission function in the vicinity of the left Fermi
level. Thus, Eq. (\ref{S2}) can be further simplified as
$S=-\frac{\pi ^2k_B^2 T}{3e}\frac{\partial ln\tau (E)}{\partial
E}\vert _{E=\mu _L } $. The Seebeck coefficient is (negative; zero;
positive) at $V_{SD}=$ ($0.8;~-0.3;~-0.5$) V because ${\partial
ln\tau (E)} \mathord{\left/ {\vphantom {{\partial ln\tau (E)}
{\partial E}}} \right. \kern-\nulldelimiterspace} {\partial E}\vert
_{E=\mu _L } $ is ($>;\approx;~<0$). We also observe that there are more
zeroes in the Seebeck coefficient as a function of $V_{SD} $ in
Fig.~\ref{Fig2}(a). For example at $V_{SD}=0.2$ or $-0.9$~V, the
peak position of the transmission function is located in the
middle of the left and right Fermi levels such that ${\partial \tau
(E)} \mathord{\left/ {\vphantom {{\partial \tau (E)} {\partial E}}}
\right. \kern-\nulldelimiterspace} {\partial E}\vert _{E=\mu _L }
\approx -{\partial \tau (E)} \mathord{\left/ {\vphantom {{\partial
\tau (E)} {\partial E}}} \right. \kern-\nulldelimiterspace}
{\partial E}\vert _{E=\mu _R } $. Consequently, the Seebeck
coefficient is close to zero at $V_{SD} $ = 0.2 and $-0.9$~V
according to Eq.~(\ref{S2}). In the unsubstituted butanethiol
junction, the Seebeck coefficient as a function of $V_{SD} $ is
shown in the inset of Fig.~\ref{Fig2}(a). The results show that the
Seebeck coefficient remains a negative value in the whole bias
regime owing to the fact that the Fermi levels are located between the
large HOMO-LUMO gap.

In conclusion, the study investigates the thermoelectricity in the
molecular junction in both linear and nonlinear regimes. The Seebeck
coefficients are studied using first-principles calculations. The general
properties of the Seebeck effects can be very different for the
unsubstituted and amino-substituted butanethiol junction in
the two-terminal and three-terminal molecular geometries.The
research illustrates that the gate field is able to modulate and
optimize the Seebeck coefficient. Another interesting phenomenon
is the possibility to change the signs of the Seebeck coefficient by
applying the gate voltages and biases in amino-substituted butanethiol
junction. It is observed that the Seebeck coefficient is
relevant to the temperatures of the electrodes that may be applied
to the design of a molecular thermometer, and its sensibility can be
controlled by gate voltages. We also extend the investigation
of the Seebeck coefficient to molecular tunnel junction at finite biases.
As the biases increase, richer features in the Seebeck coefficient are
observed, which are closely related to the transmission functions
in the vicinity of the left and right Fermi levels.  All results show that the
molecular tunnel junction based on alkanethiols may be a promising
candidate for the design of novel thermoelectric devices in the
future.

because $\partial \ln \tau (E)/\partial E\mid _{E=\mu _{L}}$ is

The authors thank MOE ATU, NCTS and NCHC for support under Grants
NSC 97-2112-M-009-011-MY3, 097-2816-M-009-004, and 97-2120-M-009-005.

\end{document}